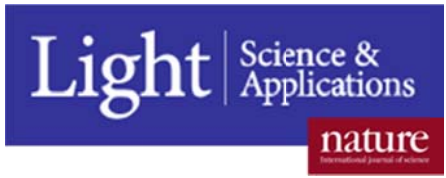

# Real-time high-resolution mid-infrared optical coherence tomography

*Niels M. Israelsen, Christian R. Petersen, Ajanta Barh, Deepak Jain, Mikkel Jensen, Günther Hannesschläger, Peter Tidemand-Lichtenberg, Christian Pedersen, Adrian Podoleanu & Ole Bang*

# Real-time High-Resolution Mid-infrared Optical Coherence Tomography

Niels M. Israelsen[1], Christian R. Petersen[1*], Ajanta Barh[2], Deepak Jain[1], Mikkel Jensen[1], Günther Hannesschläger[3], Peter Tidemand-Lichtenberg[2,5], Christian Pedersen[2,5], Adrian Podoleanu[4] and Ole Bang[1,6]

[1] DTU Fotonik, Technical University of Denmark, DK-2800 Kgs. Lyngby, Denmark
[2] DTU Fotonik, Technical University of Denmark, DK-4000 Roskilde, Denmark.
[3] Research Center for Non-Destructive Testing (RECENDT), Altenberger Straße 69, 4040 Linz, Austria.
[4] Applied Optics Group, School of Physical Sciences, University of Kent, CT2 7NH Canterbury, United Kingdoms.
[5] NLIR ApS, Hirsemarken 1, 3520 Farum, Denmark.
[6] NKT Photonics A/S, Blokken 84, DK-3460 Birkerød, Denmark

**Abstract**

The potential for improving the penetration depth of optical coherence tomography systems by using increasingly longer wavelength light sources has been known since the inception of the technique in the early 1990s. Nevertheless, the development of mid-infrared optical coherence tomography has long been challenged by the maturity and fidelity of optical components in this spectral region, resulting in slow acquisition, low sensitivity, and poor axial resolution. In this work, a mid-infrared spectral-domain optical coherence tomography system operating at 4 µm central wavelength with an axial resolution of 8.6 µm is demonstrated. The system produces 2D cross-sectional images in real-time enabled by a high-brightness 0.9-4.7 µm mid-infrared supercontinuum source with 1 MHz pulse repetition rate for illumination and broadband upconversion of more than 1 µm bandwidth from 3.58-4.63 m to 820-865 nm, where a standard 800 nm spectrometer can be used for fast detection. Images produced by the mid-infrared system are compared with those delivered by a state-of-the-art ultra-high-resolution near-infrared optical coherence tomography system operating at 1.3 µm, and the potential applications and samples suited for this technology are discussed. In doing so, the first practical mid-infrared optical coherence tomography system is demonstrated, with immediate applications in real-time non-destructive testing for the inspection of defects and thickness measurements in samples that are too highly scattering at shorter wavelengths.

**Introduction**

Optical coherence tomography (OCT) has been established as one of the most successful and significant optical techniques for biophotonics and clinical biomedical imaging, most notably within the field of ophthalmology. OCT has the ability to perform real-time, non-invasive, and non-contact measurements in reflection, providing 3D sample visualization[1,2]. Rapid advances in light sources, detectors, and components for the visible and near-infrared spectral region has enabled the development of advanced functional techniques, high-speed, and high-resolution in vivo imaging[3,4], including OCT. In recent years, there has been a growing interest in applying OCT for non-destructive testing (NDT) in cultural heritage conservation and industrial quality control to measure for example the thickness of coatings and layered materials, and to identify subsurface structures and defects[5–7]. In this regard, OCT stands out from other traditional NDT techniques, such as high-frequency laser ultrasonic (LUS) imaging, THz imaging, and micro x-ray computed tomography (µCT), each with its own advantages and disadvantages. Specifically, OCT can offer several advantages in applications prohibiting the use of µCT due to its low contrast and hazardous ionizing radiation, LUS due to the need for non-contact measurements, and THz due to poor spatial resolution and long acquisition time[7]. Furthermore, OCT is an industry-ready technology which is relatively robust and easy to apply, and can be implemented using low optical power[8]. However, the main limitation of OCT is the strong scattering of light at visible and near-IR wavelengths, which limits the penetration depth in turbid media to a few tens to hundreds of microns depending on the sample. Since scattering losses are inverse proportional to the wavelength of light relative to the size of the scattering features, it has long been known that the penetration of OCT would benefit from employing a longer center wavelength. Current state-of-the-art commercially available OCT systems for NDT operate in the 1.3 μm wavelength range, utilizing the low water absorption, and the maturity of optical fibers and components developed for telecommunications in this region. At longer wavelengths, light sources and detectors are significantly less efficient and components are less matured. In addition, water absorption is generally considered to be too strong for imaging of biological tissue and other aqueous samples in this region, and even in absence of water many materials may have significant vibrational absorption bands in this region. Therefore, the combined effect of absorption and scattering on the penetration depth makes it non-trivial to assess whether a sample would benefit from being imaged at a longer wavelength. Limited penetration is however easily confirmed at 1.3 μm where a great variety of samples have been imaged with OCT. Common for all of these investigations are their agreement with the fact that scattering is the primary obstacle for imaging deeper. Prior investigations encompassed tablet coatings[9], various polymer samples, fiber composite materials[6,10], LEDs and printed electronics[11,12], paper quality[13], as well as micro-channels in ceramics[14].

OCT studies investigating longer wavelengths in the near-IR regime at 1.7 μm and 2 μm have been reported, where increased penetration was demonstrated in composite paint samples[15,16], intralipids, rubber[17,18], and ceramic materials such as alumina and zirconia[19], although scattering was still considered to be the dominant factor in limiting penetration.

Few attempts have been made to utilize the vast potential of mid-infrared (mid-IR) OCT, such as the early proof of concept work of Colley et al.[20,21] demonstrating the first in-depth reflectivity profiles. In their work, a single reflectivity profile (A-scan) of a calcium fluoride window and a topographic image of a gold-palladium coated tissue sample were presented. These first measurements were based on quantum cascade laser (QCL) emission and cryogenically cooled mercury cadmium telluride (MCT) detection using a time-domain OCT scheme. With a center wavelength of 7 μm, Colley et al. achieved a 30 μm axial resolution with an acquisition time of 30 min for a single reflectivity profile, which was heavily affected by side lobes due to the heavily modulated spectral shape. A similar study on mid-IR OCT was reported by Varnell et al. using a superluminescent QCL centered at 5 μm with a smooth spectral shape, resulting in an improved signal-to-noise ratio (SNR), but a reduced bandwidth leading to a poor axial resolution of 50 μm[22]. Recently, Paterova et al. investigated tunable mid-IR OCT up to 3 μm using visible light detection by exploiting nonlinear interference of correlated photon pairs generated through spontaneous parametric down conversion[23]. However, due to the narrow bandwidth of their tunable source the axial resolution at 3 μm center wavelength was limited to 93 μm, while the time-domain OCT scheme chosen limited the sensitivity and acquisition speed.

In this work, a real-time mid-IR OCT system with a high axial resolution of 8.6 μm is demonstrated. This is achieved by the use of a broadband, high-brightness mid-IR supercontinuum (SC) source for illumination and a broadband frequency upconversion system for detection. Using this technique an improved penetration of the 4 μm OCT system compared to a 1.3 μm OCT system is demonstrated for the same optical power. The 1.3 μm OCT system used for comparison is a state-of-the-art spectral-domain system using a near-IR SC source with 400 nm bandwidth (for details see Supplementary materials), which was recently used in clinical skin study[24].

## Results

**Characterization of the 4 μm OCT System**

An overview of the system is shown in Fig. 1 (see Methods for details). The system consists of five modular parts: a custom mid-IR SC source based on a 1.55 μm master-oscillator power amplifier (MOPA) pump laser and a single-mode zirconium fluoride fiber, a Michelson interferometer, a scanning sample translation system, an in-house developed frequency upconversion module, and a silicon CMOS-based spectrometer. Each subsystem is connected via optical fiber to ease the coupling and alignment between subsystems. The mid-IR SC source produces a continuous spectrum from 0.9-4.7 μm and is set to operate at 1 MHz pulse repetition rate generating 40 mW of average power above 3.5 μm. The spectral components below 3.5 μm are blocked by a long-pass filter, resulting in 20 mW coupled to the sample arm of the interferometer.

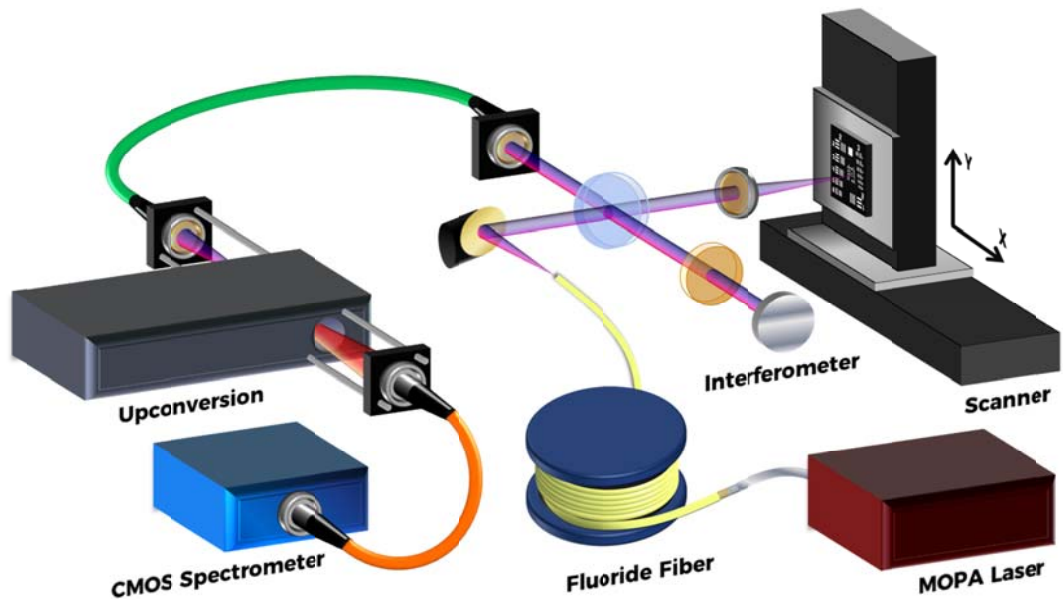

**Fig. 1 Overview of the 4 μm OCT system.** The OCT system consists of five parts that are connected via optical fiber: A broadband mid-IR SC source based on a MOPA pump laser and fluoride fiber, a free-space Michelson interferometer, a scanning x,y translation stage, a frequency upconversion module, and a silicon CMOS-based spectrometer. See Methods section for a detailed description of the set-up.

The beam is focused onto the sample using a barium fluoride ($BaF_2$) lens, and images are acquired by moving the sample using motorized translation stages. The sample and reference signals are collected in a single-mode indium fluoride fiber and relayed to the upconversion module for spectral conversion to the near-IR. The upconverted signal is then coupled to a multi-mode silica fiber and imaged onto the spectrometer to resolve the spectrum. Examples of the spectra before and after upconversion are shown in Fig. 2.a.

Here, the key technology for enabling fast and low noise detection is the broadband nonlinear frequency upconversion. Shifting the spectrum to the near-IR is critical to the performance of the system, since state-of-the-art mid-IR detectors (e.g., PbSe, InSb, MCT) suffer from intrinsic thermal background noise and low responsivity compared to their near-IR counterparts. Furthermore, due to the relative immaturity and high cost of mid-IR detectors, focal plane arrays and array spectrometers usually have a limited number of pixels available for detection, thus reducing the spectral resolution that minimizes the axial range. To alleviate these shortcomings, researchers have demonstrated frequency upconversion-based detectors that are functional at room temperature as a promising alternative to traditional direct detection schemes[25]. In this process, the mid-IR signal passes through a suitable nonlinear crystal where it mixes with a strong mixing field and as a result a near-IR sum frequency signal is generated without any loss of the information encoded in the spectral modulation of the mid-IR signal. The near-IR signal is then detected in-line by a low noise and high resolution silicon-based spectrometer. Here, the upconversion module is designed to convert a broad bandwidth of more than one micron in the mid-IR (3576-4625 nm) to a narrow band in the near-IR (820-865 nm) without any parametric tuning. To achieve this, a non-collinear angular phase-matching scheme is employed for fast parallel detection (see Methods for details). This scheme includes an MgO-doped periodically poled lithium niobate (PPLN) crystal as the nonlinear medium and a continuous wave solid state pump laser at 1064 nm with ~30 W of mixing power. The PPLN crystal is placed inside a laser cavity to gain access to the highest available power, which is directly proportional to the upconversion efficiency[26]. The module works at room temperature with a high quantum efficiency of ≥ 1% for polarized input signal over the entire targeted spectral range. The upconverted spectra of 45 nm bandwidth (corresponding to 1049 nm mid-IR bandwidth) is detected over 2286 pixels, which results in an average mid-IR spectral resolution of 0.46 nm.

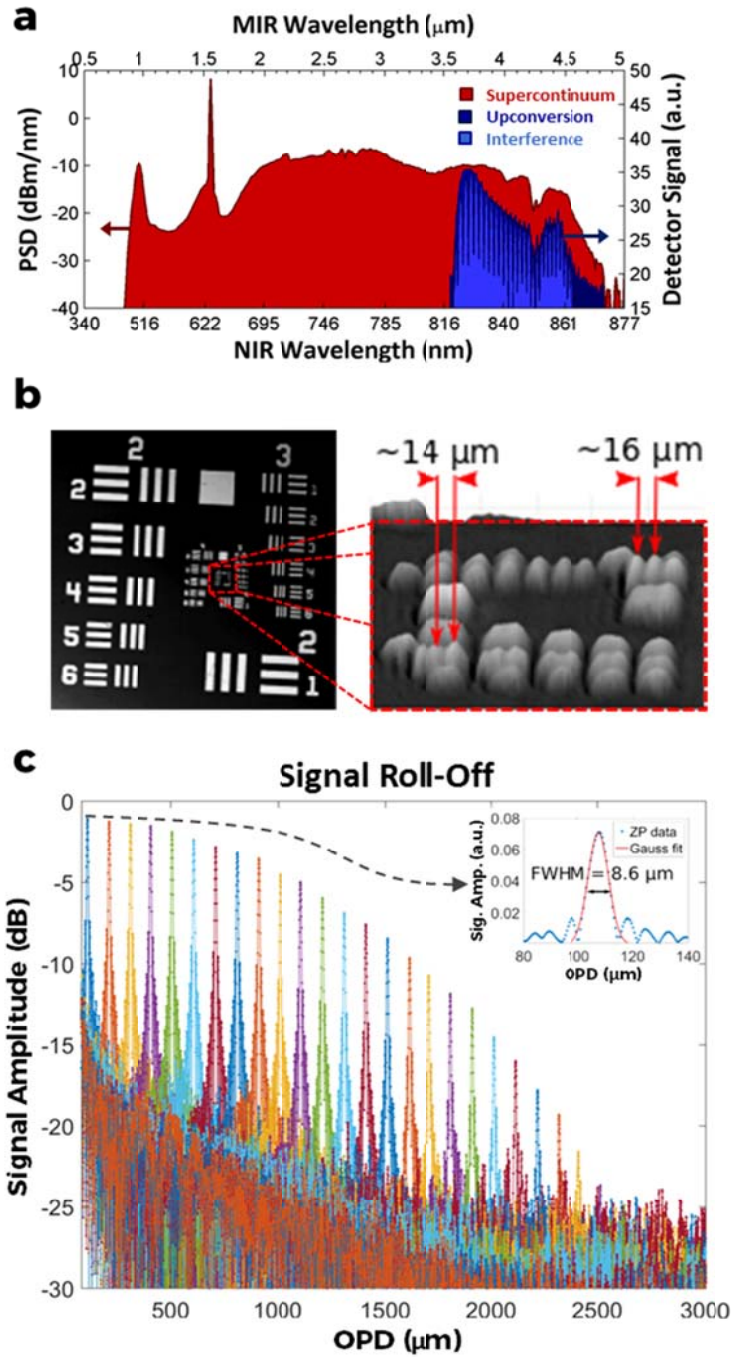

**Fig. 2 Characterization of the 4 μm OCT system. a** Superposition of SC spectra before (red) and after (dark blue) upconversion together with an example of the interference spectrum (light blue). **b** Lateral spatial resolution characterization using an USAF 1951 (left) resolution test target. The smallest resolvable features in the image (right) are elements 1 and 2 in group 6 marked by red arrows, which gives a lateral resolution of ~15 μm. **c** Sensitivity roll-off curve showing an axial range of up to 2.5 mm OPD. The inset shows a Gaussian fit of the zero-padded A-scan peak at ~100 μm OPD, giving a FWHM axial resolution of 8.6 μm.

The lateral resolution of the system is investigated by scanning a USAF 1951 resolution test target. The translation speed of the fast (horizontal) scanning axis was set to 3 mm/s, and with an integration time of 3 ms per line for maximum signal, 1000 line scans were performed in 3 s resulting in 9 µm horizontal sampling resolution. Along the vertical axis, the resolution is determined by the step size of 10 µm used to move the in the slow scanning axis of the stage was moving in 10 µm steps. From Fig. 2.b. it is shown that the system is able to resolve element 1 and 2 of group 6, which results in a lateral resolution of ~15 µm. To determine the axial resolution, the sensitivity, and the sensitivity roll-off, a plane mirror was positioned in the sample arm and the resulting interferograms were recorded for different one-way optical path difference (OPD) values by translating the mirror in the reference arm.

Fig. 2.c displays the sensitivity roll-off for 3 mm translation, which defines the effective imaging depth out of the 8.6 mm total imaging depth (given by the spectrometer resolution). From the same data an axial resolution of 8.6 µm is found at an OPD ~100 µm (see inset of Fig. 2.c), and better than 9 µm within the first 1 mm. So, even for such a long center wavelength, the axial resolution in air is similar to values reported for near-IR systems at 1.7 µm and 2 µm[15,17,18,27]. In the best case, a sensitivity of 60 dB is obtained, which is significantly lower than the 90 dB sensitivity of the 1.3 µm OCT system used for the near-IR comparison (see Supplementary section IV for details).

**Mid-IR OCT in comparison to near-IR OCT: Proof-of-principle**

As an example on the improved penetration of the mid-IR OCT system for high-resolution imaging of subsurface features, the experiments of Su et al.[19] are replicated. They demonstrated that features scanned through a ceramic plates, that were not visible using 1.3 µm OCT, became resolvable by employing 1.7 µm central wavelength. Supported by a series of Monte-Carlo simulations, Su et al. concluded that a 4 µm wavelength OCT system would be able to image through the machined alumina plate to reveal the backside of the stack. To test this, an identical set of ceramic samples were obtained through the Research Center for Non Destructive Testing (RECENDT) in Linz, Austria. A schematic of the ceramic stack is illustrated in Fig. 3.a. The stack consists of three layers of ceramic plates (C1-C3), where C1 is a 375 µm thick zirconia plate, C2 is a 475 µm thick alumina plate with up to 60 µm deep trenches and squares of varying width between 5-300 µm laser-machined into the surface, and C3 is a 300 µm alumina plate. Due to the strong scattering of alumina at 1.3 µm the near-IR system was not able to penetrate the alumina plates, so for comparing OCT at 1.3 µm and 4 µm the sample were imaged first from the top passing through the zirconia plate C1. Fig. 3.b shows A-scans obtained using the two different systems, which demonstrate that both systems detect the first interface between air and C1 at ~ 190 µm OPD and the second interface between C1 and C2 at ~1000 µm OPD. Dividing the difference in OPD with the refractive index of zirconia (2.138 at 4 µm)[19] yields a measured thickness of 388 µm. This is different by 13 µm from the 375 µm thickness that was measured mechanically using a digital thickness gauge with an accuracy of 5 µm, which is acceptable given the ~9 µm axial resolution in air. The accuracy of the OCT-based thickness measurement was later verified using a silicon wafer (see Fig. 5). Fig. 3.c shows the corresponding B-scans obtained using the two OCT systems. By comparing the two images it is clear that the mid-IR signal penetrates deeper into the sample and is able to resolve the boundary between C1 and C2 with higher contrast, as well as the boundary between C2 and C3. The near-IR OCT system provides only a weak signal from the C1-C2 boundary despite the much higher sensitivity.

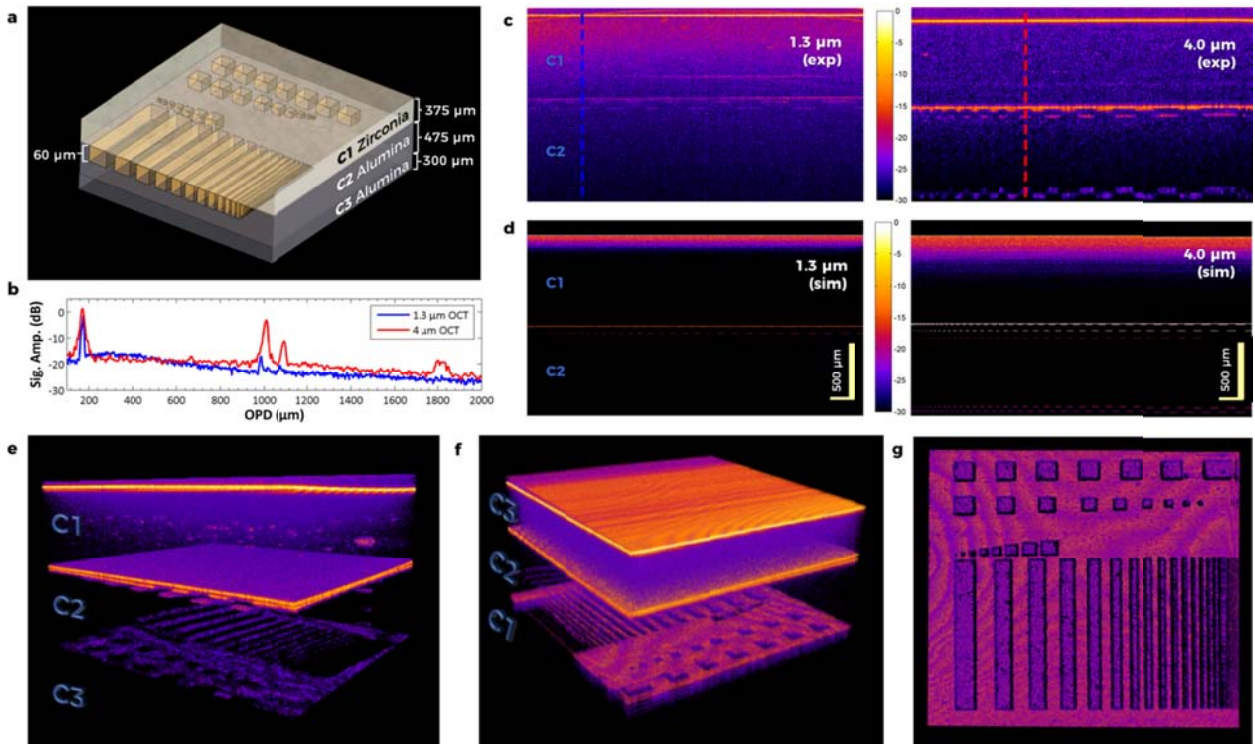

**Fig. 3 Proof of principle OCT imaging of a laser-machined ceramic stack. a** Schematic of the sample consisting of three layers of ceramics (C1-C3), where C1 is a 375 µm thick zirconia plate, C2 is a 475 µm thick alumina plate with up to 60 µm deep features of varying width between 5-300 µm laser inscribed onto the surface, and C3 is a 300 µm alumina plate. **b** Average of ten representative A-scans for the two OCT systems, as measured from the top of the sample (C1→C2→C3). **c** Comparison between B-scans using the 1.3 µm OCT (left) and 4 µm OCT systems (right). Dashed lines illustrate the position of the A-scans shown in b. **d** Monte-Carlo simulations of experiments in c. The scale bar is 500 µm in air. **e,f** 3D 4 µm OCT volume visualizations of the sample illuminating from the top (C1→C2→C3) and the bottom (C3→C2→C1), respectively. **g** En-face view of the microstructures imaged through 775 µm of alumina.

This result is in agreement with the findings of Su et al., and to further support it a series of Monte-Carlo simulations were performed (see Methods for details). The simulations, shown in Fig. 3.d, qualitatively confirm the improved visualization of in-depth interfaces in the 4 µm OCT images (see also Supplementary video I). Fig. 3.e shows a 3D visualization of the sample over a 5.6 mm x 7.3 mm scanned area, which shows strong localized scattering centers from within the porous C1 layer and a clear identification of the laser-machined structures below the C1-C2 interface (see also Supplementary videos II and III).

To further test the penetration capability of the 4 µm system, the stack was imaged from the bottom through the C3 alumina plate. Fig. 3.f shows a 3D scan of this configuration, which surprisingly reveals a significantly improved signal from the backside C2-C1 interface despite having travelled through 775 µm of alumina. As seen in Fig. 3.g the microstructures are clearly resolved, revealing surface roughness from inside the machined areas (see Supplementary Fig. S1). This demonstrates the capability of the mid IR OCT system to perform high-resolution imaging in scattering media. Due to the high spatial sampling resolution and maximum integration time used, the 800 x 730 x 2048 pixel volume took 36.5 min to acquire.

**Multiple Scattering, Dispersion, and Non-uniform Samples**
In the former experiments the reduced scattering at longer wavelengths resulted in improving the imaging depth in a semi-transparent sample. However, for very highly scattering materials, OCT imaging in the near-IR is severely impaired not only by the reduced penetration depth, but also by scrambling of the spatial image information through the effect of multiple scattering. To illustrate this, a sample consisting of a highly scattering titanium dioxide ($TiO_2$) film of varying thickness applied to a cellulose acetate foil of ~62 µm thickness was imaged with the near- and mid-IR OCT systems. $TiO_2$ is a common white pigment used in a wide variety of paints and excipients in e.g. pharmaceutical tablets[7]. Fig. 4 shows a photograph of the sample with annotations to indicate the scanned regions P1-P5, where the thickness of the $TiO_2$ layers increases from P1 to P5. The corresponding B-scans for the two systems are presented in adjacent columns for direct comparison. P1 corresponds to a B-scan of the acetate foil only, while the following images (P2-P5) portray the effect of the increased thickness of the $TiO_2$ layer imaged by the two systems. It is clear from the B-scans using the 1.3 µm OCT system that a band of multiple scattered signals increasing in thickness from P2 to P5 eventually obscures the acetate-air interface in P4 and P5, extending the apparent thickness of the $TiO_2$ layer beyond its physical extension (mirage effect). However, in the images acquired using the 4 µm OCT system the integrity of the spatial information is retained and the thickness of the $TiO_2$ layer can be evaluated. Sectioning the structure along P3, the backside of the acetate foil (n=1.48) is just barely resolved with 1.3 µm OCT, whilst clearly seen in the 4 µm B-scan. From the latter, the thickness of the $TiO_2$ layer (n=2.31)[28] is estimated to be just 19 µm, but even so the multiple scattered signals in the B-scan image at 1.3 µm extend all the way beyond the backside of the acetate foil. This significant reduction in multiple scattering going from near-IR to mid-IR represents the transition from the domain of strong directional Mie scattering with short photon mean-free-path to the domain of weaker uniform Rayleigh scattering having a long photon mean-free-path, compared to the wavelength of light.

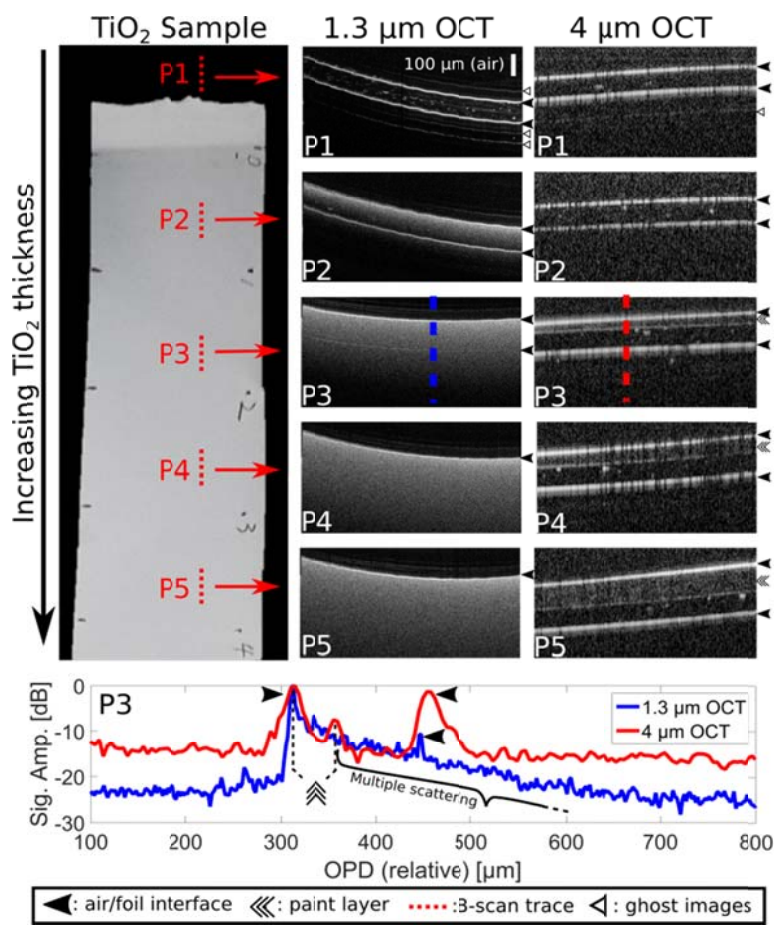

**Fig. 4 Demonstration of the reduction in multiple scattering at 4 µm. Left** Top-view photograph of the $TiO_2$ film on cellulose acetate foil with red dashed lines indicating the different B-scan sections P1-P5. **Middle** Sample B-scans at position P1-P5 using the 1.3 µm OCT system showing the detrimental effect of multiple scattering. **Right** Corresponding sample B-scans at position P1-P5 using the 4 µm OCT system showing significantly

reduced scattering. **Bottom,** Average of ten A-scans showing the traces in P3 for both OCT systems (**Symbols**. Filled arrow: air interfaces, triple arrow: $TiO_2$ layer, dashed line: B-scan traces, hollow arrow: imaging artefacts).

The mid-IR system can also have an advantage even in highly uniform and transparent media, such as germanium due to its bandgap in the near-IR, or silicon due to its lower group-velocity dispersion (GVD) at 4 µm. The effect of dispersion is to introduce a difference in delay between the short and long wavelengths of the spectrum, which in turn results in broadening of the backscattered signal. For this reason, correct depth resolved information in OCT can only be obtained after dispersion compensation[29]. To illustrate this, a 255 µm (+/-5 µm) thick silicon wafer was imaged using the two systems having a GVD of 1576 $fs^2$/mm and 385 $fs^2$/mm at 1.3 µm and 4 µm, respectively[30]. At 1.3 µm the strong dispersion results in significant broadening of the reflection peak of the wafer from 5 µm to 18 µm full width at half maximum (FWHM), whereas at 4 µm the peak broadened significantly less from 12 µm to 17 µm FWHM, as seen in Fig. 5. As a result, the image distortion due to dispersion is less pronounced in the 4 µm OCT system, which could be useful for characterization of silicon-based devices, such as micro electro-mechanical systems (MEMS), solar cells, and waveguides. Thicknesses measured with the 4 µm and the 1.3 µm OCT systems are 258.2 +/-1.2 µm and 258.6 +/- 0.56 µm respectively, when accounting for the refractive indices and with the uncertainty given by the digital sampling.

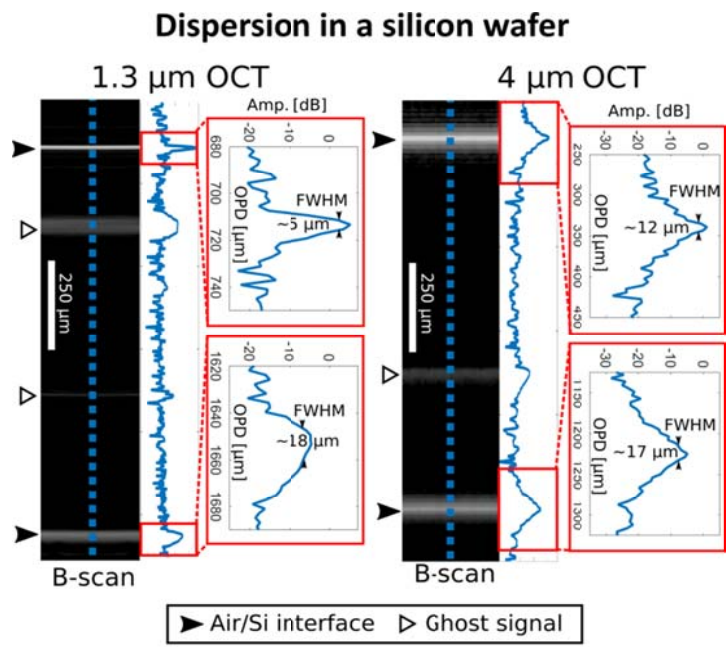

**Fig. 5 Demonstration of the reduction in dispersion at 4 µm.** B-scans of a 255 µm thick silicon wafer using the 1.3 µm (left) and 4 µm (right) OCT systems showing broadening of the A-scan peak from 5 µm to 18 µm and from 12 µm to 17 µm, respectively. Note that the peaks at the air-silicon interfaces deviate from the best case resolution due to spectral apodization.

Finally, to demonstrate 3D imaging of more complex non-uniform structures, an EMV chip and a near-field communication (NFC) antenna embedded in a standard credit card are imaged at 4 µm. Credit cards are commonly made from several layers of laminated polymers mixed with various dyes and additives. An overview of the scanned chip area is shown in Fig. 6.a. and Fig. 6.b in an en face and cross-sectional view, respectively (See also Supplementary video IV and Supplementary Fig. S2). Underneath the thin transparent surface, three layers of a highly scattering polymers are identified, which are seen most clearly at the edges of Fig. 6.b as bright uniform bands (P1-P3) separated by dark lines. As seen in Fig. 6.d, the polymer is so scattering in the near-IR, that even the top most layer (P1) cannot be penetrated by the 1.3 μm OCT system. On the other hand, the 4 μm OCT system is able to resolve all three polymer layers and can in some places even detect the backside of the card, which is 0.76 mm thick. Below the first scattering polymer layer the first structure to appear is the encapsulation layer protecting the embedded silicon microprocessor. The bonded wires and circuitry connecting the microprocessor to the underlying gold contact pad are also clearly visible, as seen in Fig. 6.c.

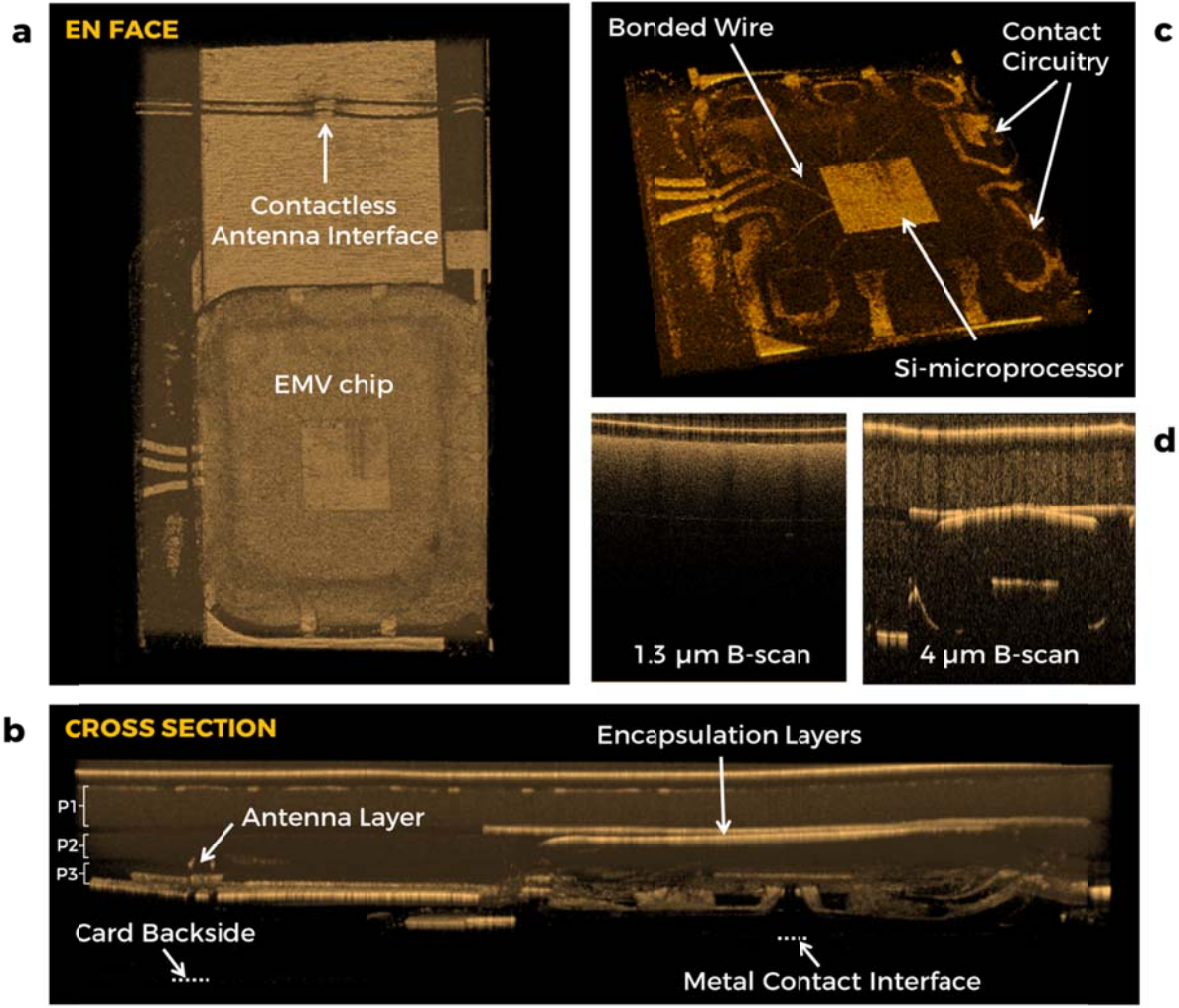

**Fig. 6 Volume 3D scan of a EMV-chip and NFC antenna embedded in a credit card. a,b** En-face and cross-sectional views of the scanned area, respectively. **c** Zoom-in on the EMV chip. **d** Comparison between B-scans obtained using the 1.3 μm OCT (left) and 4 μm OCT (right) systems.

## Discussion and conclusion

From the images of various samples presented here it would seem that the mid-IR system is superior to the near-IR in every aspect. Even with a significantly reduced sensitivity, the reduced scattering at 4 μm allows clear depth resolved imaging of interfaces from samples considered too thick for the 1.3 μm OCT system. The increased water absorption at long wavelengths naturally excludes biological samples, but otherwise the spectral region 3.6-4.6 μm is remarkably devoid of vibrational resonances, making it ideal for non-destructive testing using OCT. It is therefore expected that most materials and samples with low water content, that are currently being analyzed at 1.3 µm, but are limited to few tens to hundreds of micron penetration due to scattering, can benefit from being imaged at 4 µm. The measured axial resolution is 8.6 µm, which agrees well with the expected value for a zero-padded harmonically modulated spectrum ranging from 3576 to 4625 nm, obtained for the shortest OPDs. To improve the resolution further it is necessary to increase the bandwidth of the detected spectrum beyond the already remarkably broad >1 µm bandwidth, which is currently limited by the bandwidth of the upconversion scheme and the long-wavelength edge of the SC source. The sensitivity roll-off is currently limited by the radially varying spatial mode profile from the upconversion module, which enforces the use of a large multi-mode fiber as the limiting aperture for the spectrometer. This could be optimized if the upconverted beam could be efficiently transformed to match the fundamental mode of a single mode collecting fiber, for which the spectrometer was designed. Fulfilling this, the full image range of 8.6 mm can be exploited as the maximum signal attenuation is 15 dB over the entire imaging range for single mode coupling to the spectrometer. Approaching a centimeter in optical path length would pave the way to a whole new series of NDT applications where both imaging at macroscopic length scales along with microscopic details are required. In the end a compromise between the upconversion efficiency, the upconversion bandwidth and the extension of the output spatial mode must be achieved. To reach 90 dB sensitivity the upconversion efficiency has to be increased tenfold. In this regard, new upconversion schemes under development may allow an increase in the intra-cavity power to as much as 100-150 W, thus improving the efficiency by a factor of 3-5. Similarly, acquisition speed could be improved through increased conversion efficiency or by increasing the signal power, so as to reduce the integration time of the spectrometer.

In conclusion, fast real-time spectral-domain OCT imaging in the mid-IR was demonstrated using a supercontinuum source in combination with parallelized frequency upconversion covering more than one micron bandwidth from 3.58 to 4.63 µm. The broad bandwidth resulted in an axial resolution as high as 8.6 µm, which together with a lateral resolution of 15 µm enabled detailed imaging of microscopic structures embedded in media that are highly scattering at the more conventional 1.3 µm wavelength. The upconverted signal was measured using a Si-CMOS spectrometer acquiring data with a line rate of 0.33 kHz, which enabled real-time B-scans at mm/s speed, and C-scans at $mm^2$/min acquisition rate. To demonstrate the superiority of mid-IR OCT, comparative data has been presented using an ultra-high resolution OCT system at the more conventional 1.3 µm wavelength. To validate the results, samples were assembled similar to those used in other previous reports. Through the realization of real-time mid-IR OCT imaging this work bridges the gap between technology and practical applications of mid-IR OCT as an industry-ready tool for NDT.

## Materials and Methods

### Supercontinuum Generation

The SC source was pumped by a four-stage MOPA using an unfolded double-pass amplifier configuration based on a 1.55 μm directly modulated seed laser diode. The seed pulse duration is 1 ns, and the repetition rate is tunable between 10 kHz and 10 MHz. The seed is subjected to three stages of amplification in erbium-doped and erbium-ytterbium-doped silica fibers, which extends the spectrum to 2.2 μm by in-amplifier nonlinear broadening. In order to further push the spectrum towards longer wavelengths, the erbium fiber was spliced to ~40 cm of 10 μm core diameter thulium-doped double-clad fiber, which extended the spectrum to 2.7 μm. The thulium-doped fiber was subsequently spliced to a short piece of silica mode-field adapter fiber having a mode-field diameter of 8 μm, which provided a better match to the fluoride fiber. The mode adapter fiber is butt-coupled to a 6.5 μm core diameter single-mode ZrF4-BaF2-LaF3-AlF3-NaF (ZBLAN) fiber from FiberLabs Inc., with a short length of around 1.5 m to reduce the effect of strong multi-phonon absorption beyond 4.3 μm.

### Interferometer

The interferometer is based on a Michelson design employing a gold coated parabolic mirror collimator, a broadband $CaF_2$ wedged plate beam splitter, a $BaF_2$ plano-convex lens in the sample arm, and a $BaF_2$ window and flat silver mirror in the reference arm. The $BaF_2$ lens was chosen to minimize the effect of dispersion, while being available at a relatively short focal length of 15 mm. At 4 μm the dispersion of $BaF_2$ is relatively low at 16.4 ps nm$^{-1}$km$^{-1}$ compared to other standard lens substrates, such as $CaF_2$ (33.0), Si (-45.8), and ZnSe (-59.9), but most importantly the dispersion slope is flat from 3.5-4.5 μm (13.6-19.1 ps nm$^{-1}$km$^{-1}$)[31]. Even so, the residual dispersion from the 6.3 mm center thickness lens was roughly compensated by a 5 mm window and the remaining dispersion was compensated numerically. Coupling to the upconversion module was performed using a 6 mm focal length aspheric chalcogenide lens and a 9 μm core diameter single-mode indium fluoride patch cable.

### Upconversion

Upconversion is realized by mixing the low energy IR photons at the wavelength $\lambda_{IR}$ with pump photons at a wavelength $\lambda_P$ to generate upconverted photons of a wavelength $\lambda_{UP}$, under energy and momentum conservation:

$$\lambda_P^{-1} + \lambda_{IR}^{-1} = \lambda_{UP}^{-1}, \qquad \Delta\vec{k} = \vec{k}_{UP} - \vec{k}_P - \vec{k}_{IR}$$

where $\vec{k}$ is the wave propagation vector, and $\Delta\vec{k}$ is a measure of the phase-mismatch amongst the three interacting waves, which should ideally be zero for maximum quantum efficiency (QE). Furthermore, the QE scales linearly with the pump power and scales quadratically with the effective nonlinear coefficient ($d_{eff}$) and the length of the nonlinear medium[26]. The mid-IR OCT system is operated at a wavelength of 4 μm with more than 1 μm spectral bandwidth. Accordingly, the upconversion module was designed and optimized to upconvert the entire spectral range from 3.6-4.6 μm for fastest detection. Among various phase matching configurations, quasi phase matching in a periodically poled lithium niobate (PPLN) crystal was chosen for the broadband upconversion, owing to its design flexibility, access to a high $d_{eff}$ (14 pm/V), and optical transparency up to 5 μm[25]. The upconversion took place inside the PPLN crystal, where each wavelength was phase-matched at a different propagation angle. Thus non-collinear interaction among the three participating waves was used to phase-match over a wide spectral range (see supplementary Fig. S3.a). As the $\lambda_{UP}$ is below $\lambda_P$, by choosing the pump wavelength $\lambda_P$=1 μm, conventional Si-CMOS detection can be engaged for $\lambda_{UP}$. Here, a solid state (Nd:YVO4) continuous wave (CW) linearly polarized laser operating at 1064 nm was used as the pump. This was driven by a broad area emitting laser diode (3 W, 880 nm). The high finesse folded solid state laser cavity was formed by mirrors M1-M7 (See supplementary Fig. S4). All mirrors are HR-coated for 1064 nm and AR-coated for 700-900 nm.

Mirror M7 is based on an un-doped YAG substrate and additionally HT-coated for the 2 – 5 µm range. Mirror M6 acts as output coupler for the upconverted light. Mirrors M4 and M5 are placed in a separate compartment to filter out the fluorescence from the laser crystal and 880 nm pump laser. A 20 mm long 5 % MgO-doped PPLN crystal is used for the experiment (Covesion, AR coated for 1064 nm, 2.8 – 5.0 µm on both facets). The PPLN crystal consists of five different poling periods (Λ) ranging from 21 – 23 µm in steps of 0.5 µm. Each poled grating has a 1 mm × 1 mm aperture and is separated by 0.2 mm wide regions of un-poled material. For different values of Λ, the phase-mismatch and hence the overall upconversion spectral bandwidth varies (see supplementary Fig. S3.b). Wider bandwidth requires larger input angles for the infrared beam, which reduces the overall QE as the effective interaction length is reduced. For a best case scenario, Λ = 23 µm was considered in the set-up. A CW intracavity power of > 30 W at 1064 nm was realized with a spot size (beam radius) of 180 µm inside the PPLN crystal. The entire system is operated at room temperature. The estimated maximum QE at each mid-IR wavelength is plotted in supplementary Fig. S3.c, considering an effective interaction length of 20 mm inside the PPLN crystal. The infrared light (output of the fiber coupled 4 µm OCT signal) is collimated and then focused into the PPLN crystal using a pair of $CaF_2$ aspheric lenses (f = 50 mm, AR coated for 2 – 5 µm). The upconverted light is collimated by a silica lens (f = 75 mm, AR coated for 650 – 1050 nm). A short-pass (SP) 1000 nm and a long-pass (LP) 800 nm filter is inserted to block the leaked 1064 nm beam and 532 nm parasitic second harmonic light, respectively. A schematic representation of the radial wavelength distribution across the transverse upconverted beam profile is shown in supplementary Fig. S4. The module is able to upconvert all wavelengths in a broad mid-infrared spectral range of 3.6 – 4.8 µm to 820 – 870 nm, simultaneously.

**Detection, scanning and data processing**

After the upconversion module the near-IR light is collected by a 100 µm core multimode silica fiber guiding the light to a line scan spectrometer (Cobra UDC, Wasatch Photonics, USA) operating with a maximum line rate of 45 kHz (for a bit depth of 10). The spectral range covered wavelengths of 796 nm to 879 nm, which is sampled by 4096 pixels. To scan the sample, this is mounted on a double translation stage (2 x ILS50CC from Newport) with a maximum travel speed of 100 mm/s, a travel range of 50 mm and a stepping resolution of 1 µm. The detected raw spectra are dark signal subtracted and normalized to the reference arm signal. Pixel to wavenumber translation and interferometer dispersion compensation is achieved by exploiting phase information across the pixel array retrieved for two reference interferograms showing clear interference fringes. In this way spectral resampling is performed to linearize wavenumber sampling[32] after which a phase shift is applied for compensating the unevenly matched dispersion in the arms of the interferometer. To suppress effects stemming from the spectral envelope of the interferograms, a Hanning spectral filter is applied to the spectral region of the interferometric signals. Finally a fast Fourier transform (FFT) is applied to generate a reflectivity profile, i.e. an A-scan. A compromise between signal strength and acquisition time is made that leads to an A-scan acquisition time of 3 ms. To build B-scans (2D images), the horizontal stage (X) is programmed to move continuously over a specified distance, achieving a 500 line B-scan in 1.5 seconds. 3D scans are built by stepping the vertical (Y) stage at a proportionate slower rate to assemble multiple B-scans.

**Monte-Carlo Simulations**

The Monte Carlo (MC) simulations are performed using the open source software MCX[33]. The simulated domain consists of 1400x9x230 uniform voxels in XYZ with a size of 5x5x5 µm³ each, corresponding to a total size of 7000x45x1150 µm³. The structure is considered uniform along the Y direction, similar to what is shown in the ridge/valley part of Fig 3.a. The slab thicknesses and the valley depths in the simulation are identical to the real ones, but the valley widths in the simulation increase along the horizontal direction from 5 µm at one side of the domain to 180 µm to the other side in steps of 5 µm for the sake of simplicity. The optical properties of zirconia and alumina used in the MC simulations are taken from Su et al.[19]. The MC software launches a number of photons into the defined structure and tracks their path taking into account the refractive index, absorption, scattering, and scattering anisotropy using a random number generator to emulate scattering statistics. The simulated source uses a beam with a Gaussian transverse distribution with a constant width (5 µm for 1.3 µm center wavelength, and 15 µm for 4 µm center wavelength), due to Monte Carlo simulations being pure geometrical optics, which hinders the use of a diverging Gaussian beam. The OCT signal in a single A-scan is extracted from the MC simulation by, at each time step, summing the flux of photons that return to the top layer of the simulation within an angle corresponding to the NA of the system, which is 0.1 for both 1.3 µm and 4µm. The simulations thus emulate a time-domain OCT implementation. The simulations ran for T = 14.4 ps, in steps of dt = 24 fs, giving a digital sampling of dz =(c dt) ∕ 2 = 3.6 µm in air. Due to the lack of wave nature of light in the MC simulations, both the axial and lateral resolutions are limited by the digital sampling only, which is unphysical of course, but accepted here because the investigation focuses on the comparison of penetration depth. The B-scans are constructed by collecting A-scans obtained by moving the source 10 µm laterally, which gives 700 A-scans in a B-scan. Each A-scan simulation is performed using a different seed for the random number generator. To make the raw B-scans represent their experimental counterparts, they are corrected for roll-off by multiplying each A-scan with the measured roll-off curve, as well as corrected for the reduction in signal strength caused by the divergence of the Gaussian beam. This correction is introduced by multiplying the signal at a given depth, z, by the overlap integral, C, between the fundamental mode of the fiber that collects the light and the transverse distribution of the light reflected at z. This overlap integral is approximated here by the overlap integral between the Gaussian distribution of the propagating beam at the focal point $U(x,y,z_F)$ and the transverse distribution of light that is reflected at another point $z > z_F$. Due to the double pass of light the second distribution becomes Gaussian like the input beam at a distance $2(z-z_F)$ from the focal point:

$$C \approx \frac{\left(\iint_{-\infty}^{\infty} U(x,y,2z-z_F)U(x,y,z_F)dxdy\right)^2}{\iint_{-\infty}^{\infty} U(x,y,2z-z_F)^2 dxdy \iint_{-\infty}^{\infty} U(x,y,z_F)^2 dxdy} = \frac{4w_0^4 \pi^2 (4(z-z_F)^2 \lambda^2 + \pi^2 w_0^4)}{(4(z-z_F)^2 \lambda^2 + 2\pi^2 w_0^4)^2}$$

where $U(x,y,z)$ represents a Gaussian beam with $e^{-2}$ width $w_0$. The focal point, $z_F$, is placed 50 µm below the top surface of C1 to emulate the experimental conditions. The exact parameters and options inputs to the MCX software as well as the files specifying the spatial domain and the overlap integrals are freely available from the authors upon request.

## Acknowledgements

The authors acknowledge financial support from Innovation Fund Denmark through the ShapeOCT grant No. 4107-00011A. AP acknowledges the NIHR Biomedical Research Centre at Moorfields Eye Hospital NHS Foundation Trust and the UCL Institute of Ophthalmology and the Royal Society Wolfson Research Merit Award. D. J. acknowledges the support from H. C. Ørsted COFUNDED Marie-Curie action fellowship, and financial support from H. C. Ørsted for running cost. The authors would also like to thank the following: Dr. Lasse Høgstedt of NLIR, Nonlinear Infrared Sensors, for helpful discussions regarding the upconversion system, Andreas Buchsbaum, former employee at RECENDT for help in providing test samples, Michael Maria of Kent University for fruitful discussions regarding the OCT part, and Brian Sørensen of DTU Fotonik for assistance with the control electronics.

## Author contributions

O.B. was the principle investigator setting up the consortium and raising the funding for the project. N.M.I., C.R.P. and A.B. were the main responsible for the experimental work and manuscript writing. N.M.I. together with C.R. P., and A.P. were responsible for the 4 µm OCT system. C.R.P. together with D. J. and O.B. was responsible for the supercontinuum source. A.B. together with P.T.L. and C.P. were responsible for the upconversion detection system. N.M.I. together with A.P. were responsible for the 1.3 µm OCT system. M. J. was responsible for 1.3 µm 4 µm Monte-Carlo simulations. G.H. was responsible for supplying test samples. All authors discussed the results and contributed to the finalization of the manuscript.